\documentclass[12pt]{article}
\usepackage[superscript]{cite}
\usepackage{authblk}
\usepackage{graphics,graphicx}

\title{Fast Radio Bursts from neutron stars plunging into black holes}
\author[1]{Sudip Bhattacharyya}
\affil[1]{Department of Astronomy and Astrophysics, Tata Institute of Fundamental Research, Mumbai 400005, India; sudip@tifr.res.in}
\date{} 
\begin{document}
\maketitle


\begin{abstract}
Fast radio bursts (FRBs) are millisecond-duration intense radio flares occurring at cosmological distances.
Many models have been proposed to explain these topical astronomical events, but none has so far 
been confirmed. Here we show that a novel way involving enhanced giant radio pulses from a rapidly spun-up neutron star 
near a spinning black hole can explain the main properties of non-repeating FRBs.
Independent observations of such pulses, which are not enhanced, from some Galactic pulsars make our model reliable.
If correct, our model would imply the existence of event horizons, the Lense-Thirring effect, and
a significant spin energy extraction from a black hole. Moreover, an FRB 
would then probe the pulsar magnetosphere and its emission,
and map the strong gravity region near a black hole.
Besides, our model predicts simultaneous detections of FRBs and gravitational waves from black hole -- neutron star mergers
for fortuitously nearby FRB events.
\end{abstract}

\section{Introduction}\label{Introduction}

Fast radio bursts (FRBs) are intense millisecond-duration radio flashes, typically observed
at $\sim 1$~GHz frequency\cite{Lorimeretal2007, RaneLorimer2017, Petroffetal2016}, and are one of the most
topical astronomical phenomena of unknown origin.
The observed peak flux density ($S_{\rm peak,obs}$) is in the range of $\sim 0.02-128$~Jy, 
and when multiplied with the observed width ($W_{\rm obs}$) of the burst ($\sim 0.35-26$~ms), 
gives a fluence ($F_{\rm obs}$) in the $\sim 0.06-60$~Jy-ms range\cite{Petroffetal2016}. 
However, the most interesting property of an FRB is its high dispersion measure (DM),
which, in excess of the Galactic DM in its direction, is found to be in the range of
$\sim 140-1555$~pc cm$^{-3}$\cite{Petroffetal2016}. The implied upper limit of the luminosity distance ($D_L$) in the range
of $\sim 0.54-9.25$~Gpc\cite{Petroffetal2016}, the occurrence of FRBs at all Galactic latitudes, and 
the lack of a correlation between their DMs and Galactic latitudes\cite{RaneLorimer2017} strongly suggest that 
they occur at cosmological distances. This is confirmed for one FRB (FRB 121102), for which
a $D_L$ of 0.972 Gpc (redshift $z \approx 0.19$) has been independently estimated
\cite{Chatterjeeetal2017, Tendulkaretal2017}. These make FRBs as interesting 
as gamma-ray bursts (GRBs), and a cosmological tool, for example to study the properties of
the intergalactic medium (IGM).

A typical observed peak flux density of 1 Jy at $1.4$~GHz for a $D_L$ 
of 1 Gpc implies an instantaneous peak luminosity of $1.68\times10^{42}$~erg s$^{-1}$.
However, note that the intrinsic burst width ($W_{\rm intr}$) can be much less than the observed width $W_{\rm obs}$,
as the propagation of the radio signal broadens it. For example, while $W_{\rm obs} = 0.35$~ms, 
$S_{\rm peak,obs} = 128$~Jy and $F_{\rm obs} = 44.8$~Jy-ms for FRB 150807\cite{Petroffetal2016}, 
its $W_{\rm intr} < 40$~$\mu$s\cite{Ravi2017},
implying an intrinsic peak flux density $S_{\rm peak,intr} > 1120$~Jy. For a plausible $D_L = 0.6$~Gpc (upper limit
is 0.94 Gpc\cite{Petroffetal2016}), this implies a luminosity above $6.8\times10^{44}$~erg s$^{-1}$.
Any successful model will have to explain such a high radio luminosity with a lifetime of tens of microseconds,
that too without any significant emission in other wavelengths. This is because, to the best of our knowledge, 
no FRB source has so far been detected in other wavelengths\cite{RaneLorimer2017, Bhandarietal2017}.
Besides, some FRBs have shown structures in the flux density versus time profile\cite{Championetal2016}, while 
some have shown linear and/or circular polarization\cite{Masuietal2015, Ravietal2016, Petroffetal2016}.
In addition to these, the rate of FRBs will also have to be compared with the expected model source population. Note
that, apart from FRB 121102\cite{Spitleretal2016}, every other FRB source has shown only a single burst so far\cite{RaneLorimer2017, Bhandarietal2017}.
This indicates that either the FRB 121102 source and other FRB sources are of different classes, or they are
different evolutionary stages of the same class. We focus on the single burst sources in this paper.

It is not yet established how FRBs are generated. However, 
FRBs remind us of normal pulsars for the following reasons\cite{Cordes2016, Katz2016a, Lyutikovetal2016, Lyutikov2017}. 
(1) The FRB flashes look similar to a single pulse of pulsars.
(2) An intrinsic FRB duration of tens of $\mu$s\cite{Ravi2017} (see above) implies a $\sim 10$~km size of
the emission region, which is consistent with a neutron star size.
(3) A very high FRB brightness temperature ($\sim 10^{35}$~K) requires a coherent emission as seen from pulsars.
(4) An estimated equipartition magnetic field energy density at the source is consistent with that of pulsars\cite{Lyutikov2017}.
It is therefore not surprising that a number of proposed FRB models involve a neutron star magnetosphere.
For example, if a neutron star is born with a mass above the maximum non-spinning mass and is supported
by a high spin rate, it will eventually slow down due to magnetic braking, and will collapse into a
black hole. It has been envisaged that the magnetosphere of such a collapsed neutron star could detach, reconnect
and emit an intense electromagnetic radiation, which could be observed as an FRB\cite{FalckeRezzolla2014}.
In another model, an FRB could be generated from a binary neutron star merger due to the magnetic 
interaction between the two neutron stars\cite{Piro2012, Totani2013, Wangetal2016}. However, while such
mergers have now been confirmed to be progenitors of short GRBs\cite{Abbottetal2017},
so far no association between an FRB and a GRB has been established. 
It was also proposed that an inspiralling black hole -- neutron star binary could produce an FRB,
as the black hole, while moving through the neutron star magnetosphere, may act like 
a battery\cite{Mingarellietal2015}. 
According to yet another model, a collision of an asteroid with a neutron star could lift electrons from the 
asteroidal surface, and these accelerated electrons in the neutron star magnetosphere could cause 
an FRB\cite{Geng2015, Dai2016, Bagchi2017}. 
While these proposed mechanisms have potential to explain some observed
properties of FRBs, it is entirely unknown what types of emission they would generate. Particularly, 
due to the lack of an independent observational verification,
it is uncertain if any of them would generate the radio intensity required to explain FRBs, while
emission in other wavelengths would be too faint to detect.
For example, the above mentioned mechanism related to an inspiralling black hole -- neutron star binary
was originally proposed to emit X-rays and $\gamma$-rays\cite{McWilliams2011}.
Furthermore, while a possible radio emission related to a magnetar giant flare could 
explain some properties of FRBs\cite{Katz2016b}, such a radio emission has been ruled
out for the 2004 December outburst of the magnetar SGR 1806--20\cite{Tendulkaretal2016}.

There is another neutron star magnetospheric model of FRBs, based on giant radio pulses\cite{Cordes2016}, 
which does not have above mentioned uncertainty in radio emission.
This is because, unlike some of the proposed emissions mentioned above,
the giant pulses have actually been observed independent of FRBs.
They are regularly seen in radio from a few pulsars (e.g., Crab),
and they do not have corresponding giant pulses in other wavelengths\cite{Lyne2012}. 
Therefore, they can explain the lack of FRB counterparts in other wavelengths.
The short durations ($\sim 1-100$~$\mu$s), flux density profile structures and both
linear and circular polarizations of giant pulses are consistent with these properties 
of FRBs\cite{Lyne2012, Hankinsetal2003, Lyutikovetal2016}.
However, the Crab giant pulses can have a peak flux density of $\sim$ MJy\cite{Lyne2012, Hankinsetal2003, Lyutikov2017}
for a distance of 2 kpc, which implies a flux density of a few mJy for a distance of 1 Gpc.
Therefore, the power of giant pulses have to be enhanced, if they have to explain FRB flux densities.
The giant pulses are powered by the spin-down power of pulsars, and the instantaneous radio efficiency of these pulses
can be a few percent of the spin-down power\cite{Lyutikovetal2016}. The pulsar spin-down power is 
proportional to $B^2 R_{\rm NS}^6 \nu^4$\cite{Lyne2012}, where $B$, $R_{\rm NS}$ and $\nu$ are neutron star
surface magnetic field, radius and spin frequency respectively.
So the giant pulses from a very young pulsar, which may have a Crab-like $B$ value but a 
much higher spin frequency, could explain an observed FRB flux density\cite{Lyutikovetal2016}.

There are, however, the following inadequacies in this giant pulse based FRB model.
(1) While this model could possibly explain the FRBs with relatively low peak flux densities,
more intense FRBs would require unusually high $\nu$ values of young pulsars.
(2) Repeated bursts may be expected for this model, which is not consistent with non-repeating FRBs,
which we focus on here.
(3) Even if a young pulsar has a very high $\nu$ value, which is required to explain a more intense FRB,
it should happen at a very initial phase after its birth, because the pulsar would quickly spin down.
Therefore, an FRB source is expected to be associated with a supernova.
But such a supernova has not been detected so far. For example, 
the follow-up observations with 12 telescopes (from X-ray to radio) after the real-time detection of the 
FRB 140514 could not detect any multi-wavelength variable afterglow expected from a supernova\cite{RaneLorimer2017}.
Besides, no afterglow was found with the follow-up observations after the real-time detection of
any of four other FRBs\cite{Bhandarietal2017}.
(4) Moreover, a supernova remnant should significantly contribute to the observed DM. But the cumulative
distribution of FRB DM does not match well with that expected for expanding supernova remnant shells\cite{Katz2016a}.
In this paper, we propose a novel way to make the giant pulse based explanation of FRBs viable.

\section{Results}\label{Results}

As indicated above, we need a neutron star with a Crab-like high $B$ value ($\sim 10^{12}-10^{13}$~G) and spinning with
a $\nu \sim$ a few hundred Hz to $1$~kHz to power an enhanced giant pulse, which is required to explain intense FRBs
within the ambit of the giant pulse model.
However, such $B-\nu$ combinations cannot be easily attained, because of a large spin-down 
torque ($\propto B^2 \nu^3$) exerted on the star\cite{Lyne2012}.
For example, the only known neutron stars, which perhaps typically spin faster than young neutron stars, are
millisecond pulsars\cite{Bhattacharya1991}. They spin up to several hundreds of Hz by 
the accretion in low-mass X-ray binaries. But these pulsars have much lower $B$ values ($\sim 10^8-10^9$~G),
and hence cannot produce giant pulses of enhanced powers.
Here we explore a new way to significantly spin up a high-$B$ neutron star.

A recent paper\cite{Chakrabortyetal2017} has discussed in detail that the spin axis of a gyroscope in a Kerr spacetime
precesses. When the gyro has a non-zero angular velocity $\Omega$, i.e., it is generally in an orbit around a spinning
black hole represented by a Kerr spacetime, its spin-precession is caused by the Lense-Thirring (LT) effect and some
other effects. The expression of this spin-precession frequency as a function of the 
radial distance $r$ from the black hole centre 
is given by equations (19) and (51) of Chakraborty et al.\cite{Chakrabortyetal2017}.
Note that, for simplicity, we will consider only the spin equatorial plane (i.e., $\theta = 90^\circ$) of the black hole 
in this paper. In the above mentioned spin-precession frequency formula,
$q$ ($= (\Omega-\Omega_{-})/(\Omega_{+}-\Omega_{-})$) is used to conveniently parameterize $\Omega$, where
$\Omega_{-}$ and $\Omega_{+}$ (given by equation (26) of Chakraborty et al.\cite{Chakrabortyetal2017}) 
are non-inclusive lower and upper limits of $\Omega$ respectively. Note that $q = 0.5$ implies a non-rotating gyro
with respect to the local spacetime geometry. To make it clear with an example, if we consider two circular orbits
of the gyro around and far from a black hole, where one orbit is corotating and another is counter-rotating relative to 
the black hole spin, the former orbit has $0.5 < q < 1$ and the latter one has $0 < q < 0.5$.
Chakraborty et al.\cite{Chakrabortyetal2017} reported that as the gyro approaches the black hole, its
spin-precession frequency rises rapidly close to the event horizon, and diverges at the horizon.

In reality, in a black hole -- neutron star binary system, a spinning neutron star or a pulsar can be considered 
as a gyro rotating around a black hole\cite{Kocherlakotaetal2017}. In this paper, we, for the first time
to the best of our knowledge, propose
such neutron stars with precessing spin axes as FRB sources. Suppose, such a neutron star with a slow spin approaches a black hole.
Its spin-precession frequency should be very low at a large distance from the black hole. But as the star comes 
very close to the black hole, the spin-precession frequency can become much greater than the spin frequency.
At this time, the latter frequency may be neglected compared to the former, and the neutron star magnetosphere
can be considered to be rotating around the spin-precession axis. From this time, such a rotation and the 
corresponding spin-precession frequency will primarily power the pulsar emission, and therefore we will henceforth 
call this spin-precession frequency simply the `spin frequency' ($\nu$) of the neutron star. A high 
$\nu$-value can be attained close to the black hole, even for a high $B$-value, because the pulsar angular 
momentum will increase at the expense of a small fraction of the black hole angular momentum. 
Note that, while the angular momentum of a black hole of mass $M = 12 M_\odot$ and spin parameter $a/M = 0.75$
is $\sim 10^{51}$~g cm$^2$ s$^{-1}$, that for a fast-spinning neutron star is typically 
$\sim 10^{49}$~g cm$^2$ s$^{-1}$\cite{Bhattacharyyaetal2017}.
Therefore, the $\nu$ value could increase up to the break-up 
limit\cite{Cooketal1994, Bhattacharyyaetal2016} of the pulsar, which, depending
on its mass and equation of state, could be around $1500$~Hz. A neutron star merging with a black hole
is expected to be middle-aged or somewhat old. 
This is because a $12 M_\odot$ mass black hole and a $1.4 M_\odot$ mass neutron star would merge in about 
$10^7$ yr due to gravitational radiation, if their initial distance $r_{\rm ini}$ is $\sim 10^5$~km.
Here we use the ($5c^5r_{\rm ini}^4$)/($256G^3M_{\rm tot}^2M_{\rm red}$) expression of the merging timescale,
where $M_{\rm tot}$ and $M_{\rm red}$ are total and reduced masses of the binary system respectively\cite{Shapiro1983}.
However, for such an age of the neutron star, its $B$ value
can be in the range of $\sim 10^{12}-10^{13}$~G, as the magnetic field of such a non-accreting star
is not expected to significantly decay before $\sim 10^8$~yr\cite{Mukherjee1997}. Thus in a 
black hole -- neutron star binary system, the neutron star can have high-$B$ and high-$\nu$ combination, 
when it is very close to the black hole. Such a rejuvenated pulsar could emit enhanced giant pulses.

\begin{figure}[h]
\begin{center}
\includegraphics*[width=10cm,angle=0]{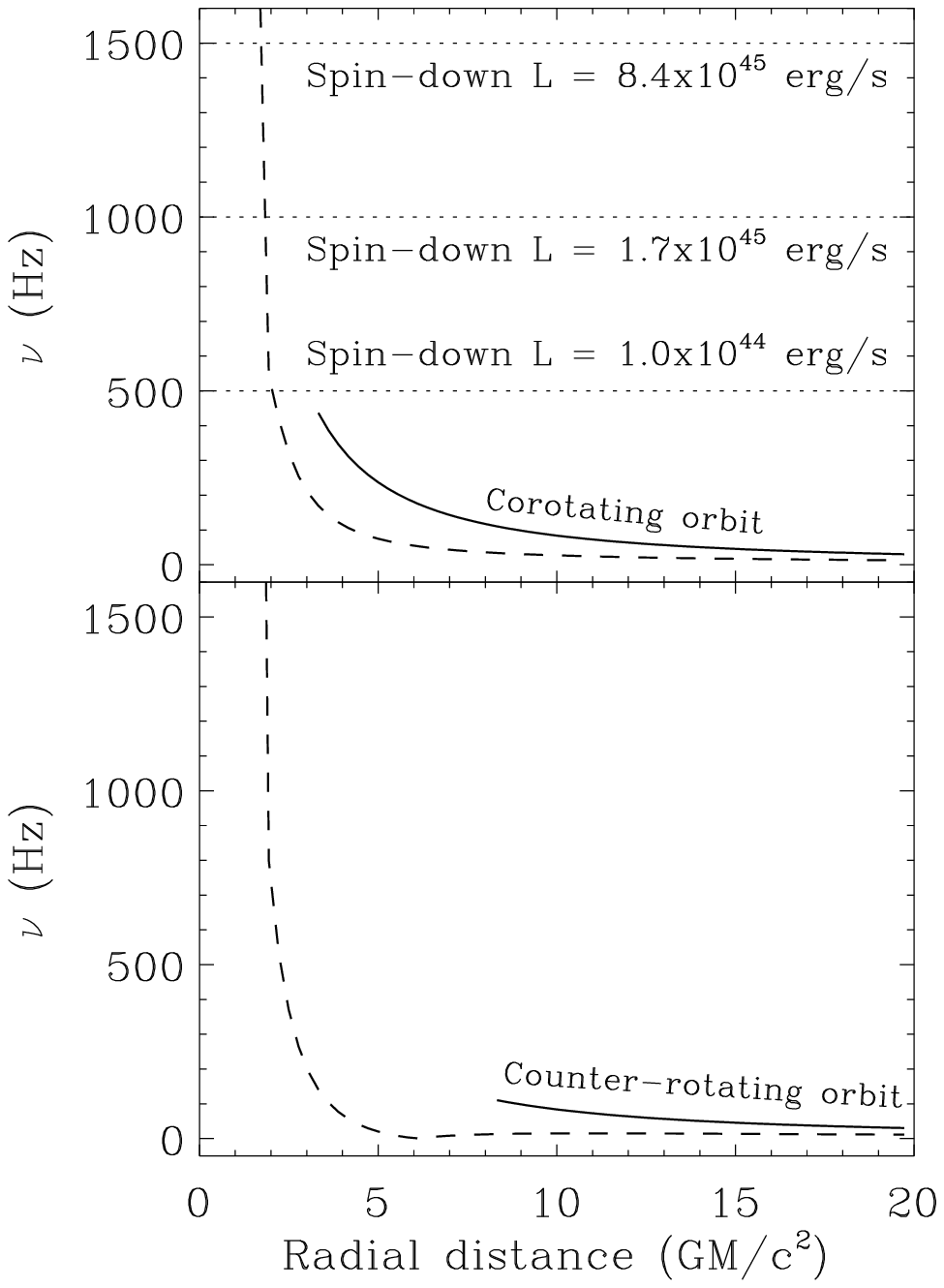}
\end{center}
\caption{Evolution of spin-preseccion frequency ($\nu$), which is the `spin frequency' for all practical purposes 
	(see text), of a neutron star rotating around a black hole of mass
	$M = 12 M_\odot$ and spin parameter $a/M = 0.75$, as the star approaches the black hole.
	The neutron star spin-down luminosities ($\propto \nu^4$) for three $\nu$ values (500 Hz, 1000 Hz 
	and 1500 Hz) due to the stellar magnetosphere rotating around the spin-preseccion axis are shown
	for a stellar surface magnetic field $B = 3.79\times10^{12}$~G (the Crab value) and a stellar 
	radius $R_{\rm NS} = 12$~km.
	{\it Upper panel:} The neutron star first moves in a corotating shrinking circular orbit ($\nu$ 
	evolution is shown by the solid curve, extended up to $r_{\rm ISCO}$), and then, as it plunges into 
	the black hole inside $r_{\rm ISCO}$, its $\nu$ value rapidly increases, as indicated by the dashed
	curve (for $q = 0.55$; see text).
	{\it Lower panel:} Similar to the {\it upper panel}, but for a counter-rotating shrinking circular orbit
	(the solid curve), and the dashed curve is for $q = 0.45$.
\label{fig1}}
\end{figure}

Now, using Fig.~\ref{fig1}, we discuss what could happen when a neutron star approaches a black hole.
For the purpose of demonstration, we consider the following reasonable parameter values: black hole 
of mass $M = 12 M_\odot$ and spin parameter $a/M = 0.75$, and neutron star radius $R_{\rm NS} = 12$~km.
For such a parameter combination, the neutron star is not expected to be tidally disrupted by the black hole,
which we estimate from the simple Newtonian fomula of the disruption radius\cite{StoneLoeb2012}:
$R_{\rm NS}(M/1.4M_\odot)^{1/3}$, and also based on more advanced results mentioned in 
Shibata \& Taniguchi\cite{Shibata2011}.
The neutron star should follow a circular orbit up to the innermost stable circular orbit (ISCO) 
radius $r_{\rm ISCO}$, as any elliptic orbit would be circularized due to gravitational radiation\cite{Shapiro1983}.
Therefore, up to $r_{\rm ISCO}$ the neutron star would
rotate with an angular velocity $\Omega$ given by equation (36) of Chakraborty et al.\cite{Chakrabortyetal2017}.
In Fig.~\ref{fig1}, we consider two cases: corotating (upper panel) and counter-rotating (lower panel) 
neutron star with respect to the black hole spin.
In the former case, the neutron star `spin frequency' $\nu$ 
will evolve tracking the solid curve, and at $r = r_{\rm ISCO}$, $q \approx 0.62$. From this point,
the neutron star will plunge into the black hole, and as it will quickly spiral in, it will lose angular
momentum via gravitational radiation. Therefore, while we do not attempt to explore the possible paths of such a 
plunging neutron star, we expect that the value of $q$ would decrease and would approach the non-rotating value 0.5.
We, therefore, include the $\nu$ profile (dashed curve) for $q = 0.55$ as an example, 
which shows that the $\nu$-value rapidly rises near the event horizon. 
This suggests that the $\nu$-value of the neutron star would rapidly increase in the plunging phase.
When $\nu$ exceeds the break-up frequency very close to the horizon, the
neutron star may be disrupted by spin-precession\cite{Kocherlakotaetal2017}.
Since this would happen very close to the black hole,
the debris due to such a disruption may be quickly and almost entirely swallowed by the 
black hole without causing any significant electromagnetic radiation\cite{Kocherlakotaetal2017, Shibata2011}.
This would be consistent with the lack of multi-wavelength detection of FRBs. 
Note that, for $\nu = 1500$~Hz and a Crab-like $B$ value, 
the spin-down luminosity is $8.4\times10^{45}$~erg s$^{-1}$ (see Fig.~\ref{fig1}), which can easily 
explain the radio luminosity ($6.8\times10^{44}$~erg s$^{-1}$; see above) of FRB 150807, if
the instantaneous radio emission efficiency of giant pulses is a few per cent of the spin-down power,
as expected\cite{Lyutikovetal2016}. Note that this FRB 150807 luminosity is one of the higher ones among FRBs, and
even a smaller $\nu$ value would produce the same radio luminosity 
for a higher radio emission efficiency of a rejuvenated pulsar, and for higher $B$ and $R_{\rm NS}$ values.

What duration is expected for such an FRB? The intrinsic duration would generally be determined by the 
duration ($\sim 1-100~\mu$s) of a giant pulse\cite{Hankinsetal2003}. This matches with some of the 
observed intrinsic durations mentioned in Ravi (2017)\cite{Ravi2017}.
However, as can be estimated from the formula mentioned above\cite{Shapiro1983},
the merger time scale from an inner orbit (say, $\sim 4-5 GM/c^2$, where the 
$\nu$-value is already as high as a few hundred Hz; upper panel of Fig.~\ref{fig1}) is about a few ms.
If during this period the extremely rapid spin-up of the rejuvenated pulsar causes two or more giant pulses,
the FRB duration would appear to be $> 1$~ms (as reported for some cases in Ravi (2017)\cite{Ravi2017})
if the burst is not resolved, or the FRB would appear as two overlapping bursts (or a multi-peaked burst)\cite{Championetal2016}, 
if the burst is partiallly resolved.

Note that the observed FRB properties would be affected by black hole properties. For example,
$\nu$-value and hence the spin-down luminosity would be higher, and the time scales would be smaller
for smaller black holes. Moreover,
for certain parameter values, a tidal disruption event could happen when the neutron star
is approaching the $r_{\rm ISCO}$, and when an FRB has already been detected (because $\nu$ has
already attained a high value). In such a case,
the FRB could be followed by an electromagnetic radiation from such a disruption event.
In case of a counter-rotating neutron star (see the lower panel of Fig.~\ref{fig1}), the results
are qualitatively similar to those for the corotating case. 
However, for the counter-rotating case, the $r_{\rm ISCO}$ is farther from the black hole, and chances of
tidal disruption is less\cite{Shibata2011}. 

We note that the rapid spin-up of a neutron star rotating around a black hole can remove the previously mentioned
inadequacies in the giant pulse based FRB model in the following way.
(1) As shown above, the instantaneous luminosity of even intrinsically very luminous FRBs can be explained.
(2) The black hole  -- neutron star merger soon after the FRB can explain the non-repeating nature of
FRBs. 
(3) Since our proposed mechanism would typically involve middle-aged or somewhat old pulsars, 
a supernova should not be associated with FRBs, which is consistent with observations as discussed above.
Finally, we note that the estimated FRB event rate ($\sim 10^3$~Gpc$^{-3}$ yr$^{-1}$)\cite{Zhang2016, Wangetal2016}
is not inconsistent with the plausible optimistic estimate rate ($\sim 10^3$~Gpc$^{-3}$ yr$^{-1}$) 
of black hole -- neutron star mergers\cite{Abadieetal2010}.

\section{Discussion}\label{Discussion}

The giant pulse based model of FRBs is uniquely reliable in the sense that we independently know the 
properties of such pulses, and, like FRBs, these pulses are seen in radio wavelengths.
Such an independent verification of radio emission is not available for many of the
other FRB models. However, some observational aspects of non-repeating FRBs could not be explained with 
this model. Here we have shown that, if a neutron star spins up near a black hole, its giant pulses 
can explain all main properties of FRBs. Note that the traditional giant pulse based model could still
possibly explain the properties of the only known repeating FRB (FRB 121102\cite{Lyutikovetal2016, Kisaka2017}).
If this is true, then the repeating and non-repeating FRBs can be caused by two different evolutionary 
stages of the same class of sources, i.e., giant pulse emitting pulsars.

The extremely rapid spin-up of a neutron star near a black hole, that we consider here, has many
challenging and complex aspects. For example, the spin-precession frequency formula used here is 
for a point gyro, and the finite dimension of the neutron star could require a modification of this formula.
Another interesting question is how the structures of the neutron star and its magnetosphere would
change due to the rapid spin-precession and the tidal force of the black hole. The natures of the proposed
black hole battery effect\cite{Mingarellietal2015} and a plausible post-merger magnetospheric reconnection 
effect\cite{FalckeRezzolla2014} could also have contributions to FRB properties. We do not attempt to address
these additional complexities here, because the aim of this paper is to introduce a novel mechanism to
explain FRBs based on giant pulses using a simple physical treatment.

If our FRB model is correct, it will have a number of implications and utilities.
For example, it might imply a submillisecond pulsar, the fastest spin-up of a neutron star
by a new mechanism, the existence of an event horizon, the LT effect, a significant spin energy extraction 
from a black hole, etc. An FRB will then be useful to probe the pulsar magnetosphere and its emission,
and to map the strong gravity region near a black hole.
A fortuitous detection of an FRB from a relatively short distance, for which gravitational waves from
a black hole -- neutron star merger can be detected, could confirm our model.

\vspace{0.5cm}
\noindent
{\bf Acknowledgements}\\
We thank Chandrachur Chakraborty, Prashant Kocherlakota, Pankaj Joshi and Alak Ray for discussions, especially related to previous joint works.

\end{document}